\documentclass[12pt]{article}
\usepackage[centertags]{amsmath}
\usepackage{amssymb,amsfonts}
\usepackage[square,comma]{natbib}
\usepackage{latexsym}
\newcommand{\be}{\begin{equation}} \newcommand{\ee}{\end{equation}}
\newcommand{\bea}{\begin{eqnarray}} \newcommand{\eea}{\end{eqnarray}}
\newcommand{\bse}{\begin{subequations}} \newcommand{\ese}{\end{subequations}}
\newcommand{\n}{\nonumber}
\begin{document}
\title{\textbf{Classes of exact Einstein-Maxwell solutions}}
\author{K. Komathiraj\thanks{Permanent
address: Department of Mathematical Sciences, South Eastern
University, Sammanthurai, Sri Lanka.}\; and S. D.
Maharaj\thanks{eMail:
\texttt{maharaj@ukzn.ac.za}}\\
Astrophysics and Cosmology Research Unit,\\ School of Mathematical
Sciences, Private Bag X54001,\\ University of KwaZulu-Natal,
Durban 4000, South Africa.}
\date{}
\maketitle
\begin{abstract}
\noindent We find new classes of exact solutions to the
Einstein-Maxwell system of equations for a charged sphere with a
particular choice of the electric field intensity and one of the
gravitational potentials. The condition of pressure isotropy is
reduced to a linear, second order differential equation which can be
solved in general. Consequently we can find exact solutions to the
Einstein-Maxwell field equations corresponding to a static
spherically symmetric gravitational potential in terms of
hypergeometric functions. It is possible to find exact solutions
which can be written explicitly in terms of elementary functions,
namely polynomials and product of polynomials and algebraic
functions. Uncharged solutions are regainable with our choice of
electric field intensity; in particular we generate the Einstein
universe for particular parameter values.

\vspace{0.5cm} \noindent \emph{Keywords}: exact solutions;
Einstein-Maxwell equations; relativistic astrophysics.
\end{abstract}

\section{Introduction}
In recent years a number of authors have found solutions to the
Einstein-Maxwell field equations for static spherically symmetric
gravitational fields with isotropic matter. These exact solutions
must match at the boundary to the unique Reissner-Nordstrom metric
which is the exterior spacetime for a spherically symmetric
charged distribution of matter. The models generated are used to
describe relativistic spheres with strong gravitational fields as
is the case in neutron stars. It is for this reason that many
investigators use a variety of techniques to attain exact
solutions. A comprehensive list of Einstein-Maxwell solutions,
satisfying a variety of criteria for physical admissability, is
provided by Ivanov \cite{Iva}. The exact solutions may be used to
study the physical features of charged spheroidal stars as
demonstrated by Komathiraj and Maharaj \cite{KoMa}, Sharma
\emph{et al} \cite{ShMuMa}, Patel and Koppar \cite{PaKo}, Patel
\emph{et al }\cite{PaTiSa}, Tikekar and Singh \cite{TiSi} and
Gupta and Kumar \cite{GuKu}. These analyses indicate that the
Einstein-Maxwell exact solutions found are relevant to the
description of dense astronomical objects. Some other individual
treatments include the Sharma \emph{et al} \cite{ShKaMu} study of
cold compact objects, the Sharma and Mukherjee \cite{ShMu}
consideration of strange matter, and the Sharma and Mukherjee
\cite{Sha} analysis of quark-diquark mixtures in equilibrium.
Charged relativistic spheres may be used to model core-envelope
stellar configuration as shown by Thomas \emph{et al}
\cite{ThRaVi}, Tikekar and Thomas \cite{TiTh}, and Paul and
Tikekar \cite{PaTi} where the core consists an isotropic fluid and
the envelope comprises an anisotropic fluid.

In order to integrate the field equations, various restrictions
have been placed by investigators on the geometry of spacetime and
the matter content. Mainly two distinct procedures have been
adopted to solve these equations for  spherically symmetric and
static manifolds. Firstly, the coupled differential equations are
solved by computation after choosing an equation of state.
Secondly, the exact Einstein-Maxwell solution can be obtained by
specifying the geometry and the form of the electromagnetic field.
We follow the latter technique in an attempt to find solutions in
term of special functions and elementary functions that are
suitable for the description of relativistic charged stars. This
approach was first used by John and Maharaj \cite{JoMa} that
yielded an uncharged star which approximates a polytrope close to
the centre. This particular exact solution was extended to a wider
class of solutions by Maharaj and Thirukkanesh \cite{MaTh} in the
presence of charge. Also Thirukkanesh and Maharaj \cite{ThMa}
found a family of Einstein-Maxwell solutions that contain the
Durgapal and Bannerji \cite{DuBa} neutron star model. Komathiraj
and Maharaj \cite{KoMa} presented a general class of
Einstein-Maxwell solutions that contain Tikekar \cite{Tik}
spheroidal stars as a special case which are physically viable
neutron star models. Hence the approach followed in this paper has
proved to be a fruitful avenue for generating new exact solutions
for describing the interior spacetimes of charged spheres.

The objective of this paper is to provide systematically a rich
family of Einstein-Maxwell solutions similar to the recent treatment
of Komathiraj and Maharaj \cite{KoMa}. In Section 2, we rewrite the
Einstein-Maxwell equations as a new set of differential equations
utilising a transformation due to Durgapal and Bannerji \cite{DuBa}.
We choose particular forms for one of the gravitational potentials
and the electric field intensity, which enables us to obtain the
condition of pressure isotropy in the remaining gravitational
potential in Section 3. This is the master equation which determines
the integrability of the system. In Section 4, we integrate the
condition of pressure isotropy, for particular parameter values,
and consequently produce Einstein-Maxwell solutions in terms of
elementary functions. We demonstrate that exact solutions to the
Einstein-Maxwell system in terms of hypergeometric functions are
possible in Section 5. In Section 6, we generate two linearly
independent classes of solutions by determining the specific
restriction on the parameters for a terminating series; the general
solution can be written explicitly in terms of elementary functions.
We demonstrate that uncharged solutions are regained in the
appropriate limit. In Section 7 we discuss the physical features,
and plot the gravitational and matter variables to show that the
model is physically acceptable. Finally in Section 8, we show that
other solutions, outside the class considered in this paper, exist
to the Einstein-Maxwell system.

\section{Field equations}
We assume that the interior of a dense compact relativistic star
should be spherically symmetric. Therefore there exists
coordinates $(t, r, \theta, \phi)$ such that the line element is
of the form \be \label{eq:1}
ds^{2}=-e^{2\nu(r)}dt^{2}+e^{2\lambda(r)}dr^{2}+r^{2}(d\theta^{2}+\sin^{2}\theta
d\phi^{2})\ee The Einstein-Maxwell field equations govern the
behaviour of the gravitational field in the presence of an
electromagnetic field. The Einstein-Maxwell system becomes
\bse\label{eq:2}\bea
\frac{1}{r^{2}}(1-e^{-2\lambda})+\frac{2\lambda^\prime}{r}e^{-2\lambda}&=&\rho+\frac{1}{2}E^{2}\\\n\\
\frac{-1}{r^{2}}(1-e^{-2\lambda})+\frac{2\nu^\prime}{r}e^{-2\lambda}&=&p-\frac{1}{2}E^{2}\\\n\\
e^{-2\lambda}\left(\nu^{\prime\prime}+{\nu^\prime}^2+\frac{\nu^\prime}{r}-\nu^\prime\lambda^\prime-\frac{\lambda^\prime}{r}\right)&=&p+\frac{1}{2}E^{2}\\\n\\
\sigma&=&\frac{1}{r^{2}}e^{-\lambda}(r^{2}E)^\prime\eea\ese for
the line element (\ref{eq:1}). The energy density $\rho$ and the
pressure $p$ are measured relative to the comoving fluid
4-velocity $u^{a}=e^{-\nu}\delta^{a}_{0}$, $E$ is the electric
field intensity, $\sigma$ is the proper charge density, and primes
denote differentiation with respect to $r$. We are utilising units
where the coupling constant $\frac{8\pi G}{c^{4}}=1$ and the speed
of light $c=1$.

A different but equivalent form of the field equations is
generated if we introduce new variables \be \label{eq:3}
A^{2}y^{2}(x)=e^{2\nu(r)},~~~Z(x)=e^{-2\lambda(r)},~~~x=Cr^{2}\ee
where $A$ and $C$ are arbitrary constants. Under the
transformation (\ref{eq:3}) due to Durgapal and Bannerji
\cite{DuBa}, the system (\ref{eq:2}) becomes
 \bse\label{eq:4} \bea
\label{eq:4a}\frac{1-Z}{x}-2\dot{Z}&=&\frac{\rho}{C}+\frac{E^{2}}{2C}\\\n\\
\label{eq:4b}4Z\frac{\dot{y}}{y}+\frac{Z-1}{x}&=&\frac{p}{C}-\frac{E^{2}}{2C}\\\n\\
\label{eq:4c}4Zx^{2}\ddot{y}+2\dot{Z}x^{2}\dot{y}+\left(\dot{Z}x-Z+1-\frac{E^{2}x}{C}\right)y&=&0\\\n\\
\label{eq:4d}\frac{\sigma^{2}}{C}&=&\frac{4Z}{x}(x\dot{E}+E)^{2}\eea\ese
where dots denote differentiation with respect to $x$. The system
of equations (\ref{eq:4}) determines the behaviour of gravity for
a charged perfect fluid. When $E=0$ we regain Einsteins equations
for a neutral fluid.  In the above we have a system of four
equations in the six unknowns $\rho, p, E, \sigma, y$ and $Z$. We
are free to specify two of the six unknowns; in this treatment we
assume forms for $Z$ and $E$. Once the metric function $Z$ and the
electric field intensity $E$ are specified then the metric
function $y$ can be found by integrating (\ref{eq:4c}) which is
then second order and linear in $y$. The remaining unknowns are
then obtained from the rest of the system. This is the approach
that we follow in this paper. Hence the differential equation
(\ref{eq:4c}) is the master equation whose integration is
necessary to determine an exact solution.
\section{Master equation}
We study a particular form of the Einstein-Maxwell system
(\ref{eq:4}) by making explicit choices for $Z$ and $E$. For the
metric function $Z$ we make the choice \be\label{eq:5}Z=\frac{(1+k
x)^{2}}{(1+x)}\ee where $k$ is a real constant. Note that the
choice (\ref{eq:5}) ensures that the metric function
$e^{2\lambda}$ is regular and finite at the centre of the sphere.
When $k=1$, in the absence of charge, we regain the Schwarzschild
interior metric. Also observe that when $k=0$ we regain the metric
function considered by Hansraj and Maharaj \cite{HaMa} which
generalises the Finch and Skea \cite{FiSk} neutron star model. We
have chosen the form (\ref{eq:5}) as it provides for a wider range
of possibilities than the solutions of Hansraj and Maharaj \cite{
HaMa}, and it does produces charged and uncharged solutions which
are necessary for a realistic model.

On substituting  (\ref{eq:5}) in (\ref{eq:4c}) we obtain
\be\label{eq:6}4(1+kx)^{2}(1+x)\ddot{y}+2(1+kx)(2k-1+kx)\dot{y}+\left[(1-k)^{2}-\frac{E^{2}(1+x)^{2}}{Cx}\right]y=0\ee
It is convenient at this point to introduce the following
transformation \be\label{eq:7}
\frac{1}{k}+x=KX,~~~\frac{1-k}{k}=K,~~~\ y(x)=Y(X)\ee This
transformation enables us to rewrite the second order differential
equation (\ref{eq:6}) in a simpler form. Under the transformation
(\ref{eq:7}), equation (\ref{eq:6}) becomes
\be\label{eq:8}4X^{2}(X-1)\frac{d^{2}Y}{dX^{2}}+2X(X-2)\frac{dY}{dX}+\left[K-\frac{E^{2}K(K+1)^{2}(X-1)^{2}}{C[K(X-1)-1]}\right]Y=0\ee
in terms of the new dependent and independent variables $Y$ and
$X$ respectively.

It is necessary to specify the electric field intensity $E$ to
integrate (\ref{eq:8}). A variety of choices for $E$ is possible
but only a few are physically reasonable which generate closed
form solutions. We can reduce (\ref{eq:8}) to simpler form if we
let
\be\label{eq:9}\frac{E^{2}}{C}=\frac{\alpha[K(X-1)-1]}{K(K+1)^{2}(X-1)^{2}}=\frac{\alpha
Kx}{(K+1)^{2}(1+x)^{2}}\ee where $\alpha$ is a constant. The form
$E^{2}$ in (\ref{eq:9}) is physically palatable because $E$
remains regular and continuous throughout the sphere.  In addition
the field intensity $E$ vanishes at the stellar centre, and has
positive values in the interior of the star for relevant choices
of the constants $\alpha$ and $K$. Upon substituting the choice
(\ref{eq:9}) in equation (\ref{eq:8}) we obtain
\be\label{eq:10}4X^{2}(X-1)\frac{d^{2}Y}{dX^{2}}+2X(X-2)\frac{dY}{dX}+(K-\alpha)Y=0\ee
which is the master equation for the system (\ref{eq:4}). When
$\alpha=0$ there is no charge. Equation (\ref{eq:10}) has to be
integrated to find an exact model for a charged sphere.

\section{Special case : elementary functions}
We can immediately integrate (\ref{eq:10}) for the special case
$K=\alpha\neq 0$. Equation (\ref{eq:10}) is separable and we
obtain the solution \be
Y(X)=c_{1}(\sqrt{X-1}-\arctan\sqrt{X-1})+c_{2}\n\ee where $c_{1}$
and $c_{2}$ are  constants of integration. In terms of the
independent variable $x$ we can write   \be
y(x)=c_{1}\left(\sqrt{\frac{1+x}{K}}-\arctan\sqrt{\frac{1+x}{K}}\right)+c_{2}\n\ee
Hence the complete solution of the Einstein-Maxwell system
(\ref{eq:4}) is then given by \bse\label{eq:11}\bea
e^{2\lambda}&=&\frac{(K+1)^{2}(1+x)}{(K+1+x)^{2}}\\\n\\
e^{2\nu}&=&A^{2}\left[c_{1}\left(\sqrt{\frac{1+x}{K}}-\arctan\sqrt{\frac{1+x}{K}}\right)+c_{2}\right]^{2}\\\n\\
\frac{\rho}{C}&=&\frac{K^{2}(6+x)-6(1+x)^{2}}{2(K+1)^{2}(1+x)^{2}}\\\n\\
\frac{p}{C}&=&\frac{2c_{1}(K+1+x)}{\sqrt{K}(K+1)^{2}\sqrt{1+x}\left[c_{1}\left(\sqrt{\frac{1+x}{K}}-\arctan\sqrt{\frac{1+x}{K}}\right)+c_{2}\right]}\n\\\n\\
&+&\frac{2(1+x)^{2}-K^{2}(2+x)}{2(K+1)^{2}(1+x)^{2}}\\\n\\
\frac{E^{2}}{C}&=& \frac{K^{2}x}{(K+1)^{2}(1+x)^{2}} \eea\ese\\
Note that the charged solution (\ref{eq:11}) does not have an
uncharged analogue as the electric field intensity $E$ cannot
vanish (expect at the centre). This effect essentially results
from our condition that $\alpha=K(\neq 0)$. This means that this
solution models a sphere that is always charged and hence cannot
attain a neutral state. A particular class in the family of
solutions found by Hansraj and Maharaj \cite{HaMa} also
demonstrates the same feature and $E\neq 0$. The model
(\ref{eq:11}) is a simple solution of the Einstein-Maxwell system
which is expressed in terms of elementary functions.
\section{General case : series solution}
With $\alpha\neq K$, equation (\ref{eq:10}) is difficult to solve.
However it can be transformed to a hypergeometric differential
equation which can be integrated using the method of Frobenius. We
now introduce a new function $U(X)$ such that \be\label{eq:12}
Y(X)=X^{a}U(X) \ee where $a$ is a constant. On substituting
(\ref{eq:12}) in (\ref{eq:10}) we obtain \be \label{eq:13}
4X^{2}(X-1)\frac{d^{2}U}{dX^{2}}+2X[(4a+1)X-2(2a+1)]\frac{dU}{dX}+[2a(2a-1)X+K-\alpha-4a^{2}]U=0\ee
We observe that there is considerable simplification if we make
the choice \be \label{eq:14} K-\alpha=4a^{2}\ee This then gives
\be\label{eq:15}2X(X-1)\frac{d^{2}U}{dX^{2}}+[(4a+1)X-2(2a+1)]\frac{dU}{dX}+a(2a-1)U=0\ee
which is a second order differential equation in terms of the new
dependent variable $U$ and independent variable $X$. When $a=0$
then $\alpha=K$ and we regain the result of Section 4. Therefore
we take $a\neq 0$ in this section to ensure that $\alpha \neq K $.

If we let $z=1-X$ then (\ref{eq:15}) becomes \be\label{eq:16}z(1-z)
\frac{d^{2}U}{dz^{2}}- \left[\left(2a + \frac12 \right)z +
\frac12\right]\frac{dU}{dz}- a\left(a-\frac12\right)U=0\ee
 The result (\ref{eq:16}) is a special case of the hypergeometric
 equation which can be solved explicitly in terms of special
 functions $U_1$ and $U_2$. These special functions are
 hypergeometric functions and are given by
 \be \label{eq:17} U_1 = F\left(a, a-\frac12, -\frac12, z\right) \ee
  and
  \be
  \label{eq:18} U_2 = z^{3/2} F\left(a+\frac32, a+1, \frac52,
 z\right) \ee
  It is now possible to write the solution of (\ref{eq:16}) explicitly
  as a series using the definitions of (\ref{eq:17}) and (\ref{eq:18}).
  With the help of (\ref{eq:7}) and (\ref{eq:12}) we obtain the
  expressions
  \bea
\label{eq:19}y_{1}(x)&=& \left(\frac{K+1+x}{K}\right)^{a}\n\\
&\times&\left[1+\sum_{i=1}^{\infty}(-1)^{i}\prod_{p=1}^{i}\frac{[(p-1)(2p+4a-3)
+a(2a-1)]}{p(2p-3)}\left(\frac{1+x}{K}\right)^{i}\right]\eea and
\bea \label{eq:20}y_{2}(x)&=&
\left(\frac{K+1+x}{K}\right)^{a}\left(\frac{1+x}{K}
\right)^{\frac{3}{2}}\n\\
&\times&\left[1+\sum_{i=1}^{\infty}(-1)^{i}\prod_{p=1}^{i}\frac{[(2p+1)(p+2a)
+a(2a-1)]}{p(2p+3)}\left(\frac{1+x}{K}\right)^{i}\right]\eea as
linearly independent solutions of (\ref{eq:6}).
 Thus the general solution to the differential equation
(\ref{eq:6}), for the choice of the electric field (\ref{eq:9}),
is given by \be \label{eq:21}y(x)=A_{1}y_{1}(x)+A_{2}y_{2}(x)\ee
where $A_{1}$ and $A_{2}$ are arbitrary constants,
$K=\frac{1-k}{k}$,~ $a^{2}=\frac{K-\alpha}{4}$, and $y_{1}$ are
$y_{2}$
 are given by (\ref{eq:19}) and (\ref{eq:20}) respectively.

 From (\ref{eq:21}) and (\ref{eq:4}) we can write the exact
 solution of the Einstein-Maxwell system in the form
\bse\label{eq:22}\bea
e^{2\lambda}&=&\frac{(K+1)^{2}(1+x)}{(K+1+x)^{2}}\\\n\\
e^{2\nu}&=&A^{2}y^{2}\\\n\\
\label{eq:23c}\frac{\rho}{C}&=&\frac{(K^{2}-1)(3+x)-x(5+3x)}{(K+1)^{2}(1+x)^{2}}
-\frac{\alpha Kx}{2(K+1)^{2}(1+x)^{2}}\\\n\\
\frac{p}{C}&=&\frac{4(K+1+x)^{2}}{(K+1)^{2}(1+x)}\frac{\dot{y}}{y}+\frac{1-K^{2}
+x}{(K+1)^{2}(1+x)}+\frac{\alpha Kx}{2(K+1)^{2}(1+x)^{2}}\\\n\\
\frac{E^{2}}{C}&=& \frac{\alpha Kx}{(K+1)^{2}(1+x)^{2}} \eea\ese
Unlike the solution presented in Section 4, the models found in
this section cannot be written in terms of elementary functions in
general as the series in (\ref{eq:17}) and (\ref{eq:18}) do not
terminate. However terminating series are possible for particular
values of $a$, which leads to elementary functions, as we show in
Section 6.

\section{Elementary functions}
The general solution (\ref{eq:21}) is given in the form of a series
and can be expressed in terms of hypergeometric functions which are
special functions. It is well known that hypergeometric functions
can be written in terms of elementary functions for particular
parameter values. This statement is also true for the solution found
in Section 5 for particular values of the parameter $a$ as the two
series terminate. Consequently two sets of general solutions in
terms of elementary functions can be found by restricting the range
of values of $a$ so that the series terminates. The elementary
functions, found in this way,  are expressible as polynomials and
product of polynomials with algebraic functions. We can express the
first category of solutions, in terms of the original variable $x$,
as
\bea\label{eq:23}y(x)&=&A_{1}\left(\frac{K}{K+1+x}\right)^{n}\sum_{i=0}^{n}\frac{(
-1)^{i-1}(2i-1)}{(2i)!(2n-2i+1)!}\left(\frac{1+x}{K}\right)^{i}\n\\
&+&A_{2}\left(\frac{K}{K+1+x}\right)^{n}\left(\frac{1+x}{K}\right)^{\frac{3}{2}}\sum_{i
=0}^{n-1}\frac{(-1)^{i}(i+1)}{(2i+3)!(2n-2i-2)!}\left(\frac{1+x}{K}\right)^{i}\eea
where  $K-\alpha=4n^{2}$. The second category of solutions is
given by
\bea\label{eq:24}y(x)&=&A_{1}\left(\frac{K}{K+1+x}\right)^{n-\frac{1}{2}}\sum_{i=0}^{n}\frac{(
-1)^{i-1}(2i-1)}{(2i)!(2n-2i)!}\left(\frac{1+x}{K}\right)^{i}\n\\
&+&A_{2}\left(\frac{K}{K+1+x}\right)^{n-\frac{1}{2}}\left(\frac{1+x}{K}\right)^{\frac{3}{2}}\sum_{i
=0}^{n-2}\frac{(-1)^{i}(i+1)}{(2i+3)!(2n-2i-3)!}\left(\frac{1+x}{K}\right)^{i}\eea
where  $K-\alpha=4n(n-1)+1$.

Therefore  two categories of solutions in terms of elementary
functions can be extracted from the general series in Section 5.
The solutions in (\ref{eq:23}) and (\ref{eq:24}) have a simple
form, and they have been expressed completely as combinations of
polynomials and algebraic functions. This has the advantage of
simplifying the investigation into the physical properties of a
dense charged star. As the metric function (\ref{eq:5}) and the
electric field intensity (\ref{eq:9}) have not been considered
before, we believe that the Einstein-Maxwell solutions found here
have not been published previously. It is interesting to observe
that our treatment has brought together the charged and uncharged
models for a relativistic star. If we set $\alpha = 0$ in the
Einstein-Maxwell solutions (\ref{eq:23}) and (\ref{eq:24}) then we
obtain the solutions for the uncharged case directly. Thus our
approach has the welcome feature of producing uncharged solutions
when $E=0$; it is possible that the uncharged solutions produced
in this procedure may be new.

We illustrate this feature with an example. We observe that  when
$K-\alpha=4(n=1)$, (\ref{eq:23}) becomes
\be\label{eq:25}y(x)=\frac{a_{1}(K+3+3x)+a_{2}(1+x)^{\frac{3}{2}}}{K+1+x}\ee
where $a_{1}$ and $a_{2}$  are constants. On substituting
(\ref{eq:25}) in (\ref{eq:22}) we obtain the general solution to
the Einstein-Maxwell system of equations as \bse\label{eq:26}\bea
e^{2\lambda}&=&\frac{(K+1)^{2}(1+x)}{(K+1+x)^{2}}\\\n\\
e^{2\nu}&=&A^{2}\left[\frac{a_{1}(K+3+3x)+a_{2}(1+x)^{\frac{3}{2}}}{K+1+x}\right]^{2}\\\n\\
\frac{\rho}{C}&=&\frac{6(K^{2}-1)+x[(K+6)(K-2)-6x]}{2(K+1)^{2}(1+x)^{2}}\\\n\\
\frac{p}{C}&=&\frac{2(K+1+x)[4a_{1}K+a_{2}\sqrt{1+x}(3K+1+x)]}{(K+1)^{2}(1+x)[a_{1}(K+3+3x)+a_{2}(1+x)^{\frac{3}{2}}]}\n\\\n\\
&+&\frac{2(1+x)(1-K^{2}+x)+K(K-4)x}{2(K+1)^{2}(1+x)^{2}}\\\n\\
\frac{E^{2}}{C}&=& \frac{K(K-4)x}{(K+1)^{2}(1+x)^{2}} \eea\ese for
our chosen parameter values. When $\alpha=0(K=4)$ the
electromagnetic field vanishes and we get
\bse\label{eq:27}\bea e^{2\lambda}&=&\frac{25(1+x)}{(5+x)^{2}}\\
e^{2\nu}&=&A^{2}\left[\frac{a_{1}(7+3x)+a_{2}(1+x)^{\frac{3}{2}}}{5+x}\right]^{2}\\
\frac{\rho}{C}&=&\frac{45+x(10-3x)}{25(1+x)^{2}}\\
\frac{p}{C}&=&\frac{a_{1}[3x(x-2)+55]+a_{2}\sqrt{1+x}[x(22+3x)+115]}{25(1+x)[a_{1}(7+3x)+a_{2}(1+x)^{\frac{3}{2}}]}\eea\ese
Thus we have generated the uncharged solution (\ref{eq:27}) from
the charged solution (\ref{eq:26}).

\section{Physical  features}
We make some brief comments relating to the physics found in this
paper. In the general solution (\ref{eq:22}), when studying models
of charged spheres, we should consider only those values of $K$ for
which the the energy density $\rho$, the pressure $p$ and the
electric field intensity $E$ are positive. Our choice of the
gravitational potential (\ref{eq:5}) is clearly positive for a wide
range of the parameter values of $K$. Since
$y(x)=A_{1}y_{1}(x)+A_{2}y_{2}(x)$ given in (\ref{eq:23}) or
(\ref{eq:24}) is well defined function on the interval $[0,d]$ where
$d=CR^{2}$ and $R$ is the stellar radius, the quantities
$\nu,~\lambda,~\rho,~p$ and $E$ are nonsingular and continuous. If
$K>1(\alpha>0)$ or $K<-1(\alpha<0)$, then it is clear from
(\ref{eq:23c}) that $\rho$ remains positive in the region \be
\frac{x(10+\alpha K+6x)}{(3+x)}<2(K^{2}-1)\n\ee for positive
constant $C$, which restricts the size of the configuration. We
require that the pressure must vanish across the boundary $r=R$
which implies that \be
4\frac{(K+1+CR^{2})}{(K+1)^{2}(1+CR^{2})}\left[\frac{\dot{y}}{y}\right]_{x=CR^{2}}
+\frac{1-K^{2}+CR^{2}}{(K+1)^{2}(1+CR^{2})}+\frac{\alpha
KCR^{2}}{2(K+1)^{2}(1+CR^{2})^{2}}=0\n\ee where $y$ is given by
(\ref{eq:23}) or (\ref{eq:24}). Essentially this places a
restriction on the constants $A_{1}$ and $A_{2}$. The interior
metric (\ref{eq:1}) must match to the exterior Reissner-Nordstrom
line element \be
ds^{2}=-\left(1-\frac{2M}{r}+\frac{Q^{2}}{r^{2}}\right)dt^{2}+\left(1-\frac{2M}{r}
+\frac{Q^{2}}{r^{2}}\right)^{-1}dr^{2}+r^{2}(d\theta^{2}+\sin^{2}\theta
d\phi^{2})\n\ee at the boundary $r=R$. This requirement implies that
\bea
1-\frac{2M}{r}+\frac{Q^{2}}{r^{2}}&=&A^{2}[A_{1}y_{1}(CR^{2})+A_{2}y_{2}(CR^{2})]^{2}\n\\\n\\
\left(1-\frac{2M}{r}+\frac{Q^{2}}{r^{2}}\right)^{-1}&=&\frac{1+CR^{2}}{(1+kCR^{2})^{2}}\n\eea
This gives the relationships between the constants
$A_{1},~A_{2},~k$(or $K$), $A$ and $C$. We must have \be
Q^{2}(R)=\frac{\alpha KC^{2}R^{6}}{(K+1)^{2}(1+CR^{2})^{2}}\n\ee
to ensure the continuity of the electric field intensity across
the boundary. This shows that continuity of the metric
coefficients and matter variables across the boundary of the star
is easily achieved. The matching condition at the boundary may
place restrictions on the metric coefficients $\nu$ and its first
derivative for uncharged matter; and the pressure may be nonzero
if there is a surface layer of charge. However there are
sufficient free parameters to satisfy the necessary condition that
arises from a particular physical model under consideration.

We are  in a position to investigate the gravitational behavior of
this model in the interior of the star for particular choices of the
parameter values in (\ref{eq:26}). The behaviour of the stellar
model is illustrated in terms of graphs of the matter variables and
the gravitational potentials. We have generated these graphs with
the
 assistance of the software package Mathematica. For simplicity
 we make the choices $K=5$, $A=C=1$, $ a_1=a_2 =1$ and $\alpha =1$,
 over the interval $0\leq r\leq 1$,
 to generate the relevant plots.
In Fig. 1 and Fig. 2 we have plotted the metric functions $e^{2\nu}$
and $e^{2\lambda}$, respectively. It can easily be seen tha the
gravitational potentials
 remain regular in the interior of the star for $0\leq r\leq 1$.
In Fig. 3 we have the behaviour of the energy density $\rho$, and
Fig. 4 gives the representation for the isotropic pressure $p$. We
observe that the energy density  and  the pressure
 are positive and monotonically decreasing functions
in the interior of the star. The electric field intensity $E^2$ is
given in Fig. 5 which is positive and monotonically increasing. Thus
the quantities $\rho$, $p$, $E$, $e^{2\nu}$ and $e^{2\lambda}$ are
continuous, regular and well behaved throughout the interior of the
star. In Fig. 6  we have plotted $\frac{dp}{d\rho}$ on the interval
$1 \leq r \leq 1$. It can be observed from Fig. 6 that the speed of
sound is always less than unity. Consequently the speed of the speed
of sound is always less than the speed of light and causality is not
violated. Therefore we have demonstrated that there exist particular
values for the parameters so that the solution (\ref{eq:26})
satisfies the requirements for a physically reasonable charged star.

\section{Discussion}
We have found new solutions (\ref{eq:11}) to the Einstein-Maxwell
system (\ref{eq:4}), by utilising the coordinate transformation
(\ref{eq:7}), that do not have an uncharged analogue. These
solutions are given in terms of elementary functions; other
solutions are possible in terms of a general series. Consequently
other new exact solutions (\ref{eq:22}) to the Einstein-Maxwell
field equations were found in terms of special functions, namely
hypergeometric functions. The electromagnetic field may vanish in
the general series solutions and we can regain uncharged solutions.
It is possible for hypergeometric functions to be expressed in terms
of elementary functions for particular parameter values. We used
this feature to find two classes of exact solutions (\ref{eq:23})
and (\ref{eq:24}) to the Einstein-Maxwell system in terms of
polynomials and product of polynomials and algebraic functions. The
simple form of the solutions found facilitate the analysis of the
physical features of a charged sphere. For particular parameter
values we showed that it is possible to model a physically
acceptable charged relativistic sphere.

We should emphasise that the solutions found in the paper depend
crucially on the transformation (\ref{eq:7}) in which $k\neq 0$
and $k\neq 1$. Consequently we cannot regain the Schwarzchild
interior metric $(k=1)$ or the family of metrics of Hansraj and
Maharaj \cite{HaMa} $(k=0)$. A different coordinate transformation
from (\ref{eq:7}), allowing for $k=0$ and $k=1$, must be utilised
to regain previously known solutions; a paper outlining this
further new class of Einstein-Maxwell solutions is under
preparation. Clearly such solutions are possible as the following
example illustrates. For the choice of metric function
(\ref{eq:5}), we can show that the system (\ref{eq:4}) admits the
particular exact
solution\bse\bea e^{2\lambda}&=&\frac{1+x}{(1+kx)^{2}}\\\n\\
e^{2\nu}&=&1\\\n\\
\rho&=&\frac{C[6(1-2k)+x(1-2k-11k^{2})-6k^{2}x^{2}]}{2(1+x)^{2}}\\\n\\
p&=&\frac{C[2(2k-1)+x(2k+3k^{2}-1)+2k^{2}x^{2}]}{2(1+x)^{2}}\\\n\\
E^{2}&=&\frac{C(1-k)^{2}x}{(1+x)^{2}}\eea\ese When $k=1$ then
$E=0$ and we have uncharged matter with the line element \be
ds^{2}=-dt^{2}+\frac{1}{1+Cr^{2}}dr^{2}+r^{2}(d\theta^{2}+\sin^{2}\theta
d\phi^{2})\ee with equation of state $\rho+3p=0$. Thus we have
regained the familiar Einstein universe.\\

\noindent{\large \bf Acknowledgements}\\

\noindent We are grateful to the referee for valuable advice.
KK
thanks the National Research Foundation and the University of
KwaZulu-Natal for financial support, and also extends his
appreciation to the South Eastern University of Sri Lanka for study
leave. SDM acknowledges that this work is based upon research
supported by the South African Research Chair Initiative of the
Department of Science and Technology and the National Research
Foundation.

\newpage

\begin{figure}[thb]
\vspace{1.5in} \includegraphics{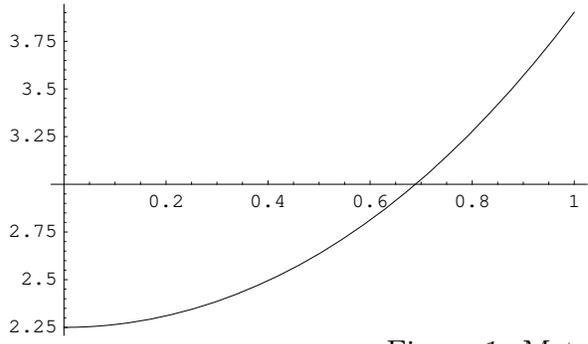} \caption{\label{plots1} Metric function
$e^{2\nu}$}
\end{figure}

\vspace{.1in}

\begin{figure}[thb]
\vspace{1.5in} \includegraphics{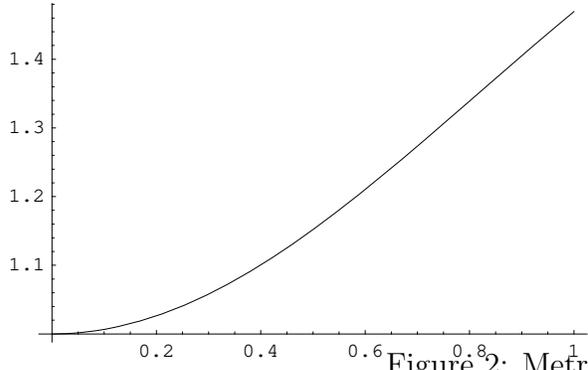} \caption{\label{plots2}Metric function
$e^{2\lambda}$}
\end{figure}

\vspace{.1in}

\begin{figure}[thb]
\vspace{1.5in} \includegraphics{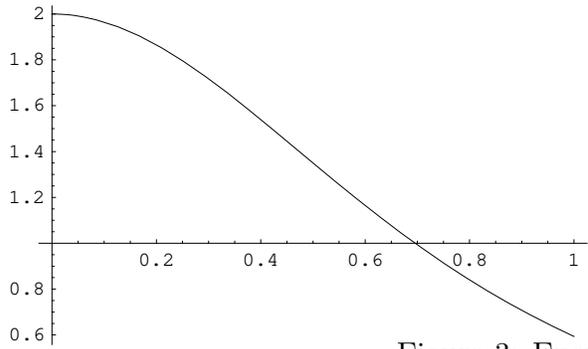} \caption{\label{plots3}Energy density
$\rho$ }
\end{figure}

\newpage

\begin{figure}[thb]
\vspace{1.5in} \includegraphics{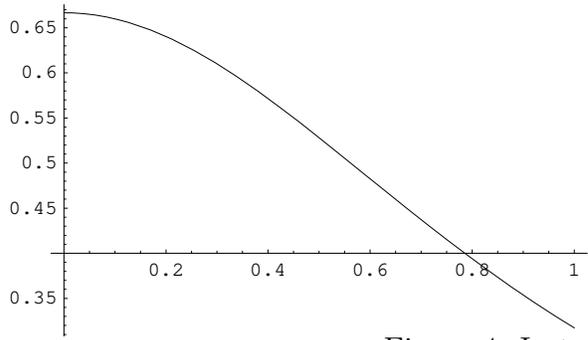} \caption{\label{plots4} Isotropic
pressure $p$}
\end{figure}

\vspace{.1in}

\begin{figure}[thb]
\vspace{1.5in} \includegraphics{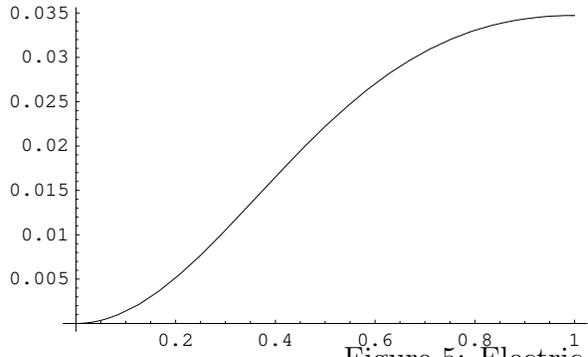} \caption{\label{plots5}Electric field
intensity $E^2$}
\end{figure}

\vspace{.1in}

\begin{figure}[thb]
\vspace{1.5in} \includegraphics{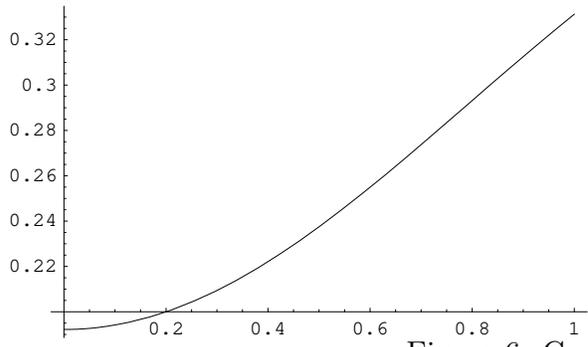} \caption{\label{plots6}  Gradient $dp/d
\rho$ }
\end{figure}

\newpage

%

%
%
%

\begin{thebibliography}{0}
\bibitem{Iva}
Ivanov, B.V.: Phys. Rev. D \textbf{65}, 104001 (2002)

\bibitem{KoMa}
Komathiraj, K., Maharaj, S.D.: J. Math. Phys., 042501 (2007)

\bibitem{ShMuMa}
Sharma, R., Mukherjee, S., Maharaj, S.D.: Gen. Relat. Gravit.
\textbf{33}, 999 (2001)

\bibitem{PaKo}
Patel, L.K., Koppar, S.K.: Aust. J. Phys. \textbf{40}, 441 (1987)

\bibitem{PaTiSa}
Patel, L.K., Tikekar, R., Sabu, M.C.: Gen. Relat. Gravit.
\textbf{29}, 489 (1997)

\bibitem{TiSi}
Tikekar, R., Singh, G.P.: Gravitation and Cosmology \textbf{4},
294 (1998)

\bibitem{GuKu}
Gupta, Y.K., Kumar, M.: Gen. Relat. Gravit. \textbf{37}, 233
(2005)

\bibitem{ShKaMu}
Sharma, R., Karmakar, S., Mukherjee, S.: Int. J. Mod. Phys. D
\textbf{15}, 405 (2006)

\bibitem{ShMu}
Sharma, R., Mukherjee, S.: Mod. Phys. Lett. A \textbf{16}, 1049
(2001)

\bibitem{Sha}
Sharma, R., Mukherjee, S.: Mod. Phys. Lett. A \textbf{17}, 2535
(2002)

\bibitem{ThRaVi}
Thomas, V.O., Ratanpal, B.S., Vinodkumar, P.C.:  Int. J. Mod.
Phys. D \textbf{14}, 85 (2005)

\bibitem{TiTh}
Tikekar, R., Thomas, V.O.: Pramana - J. Phys. \textbf{50}, 95
(1998)

\bibitem{PaTi}
Paul, B.C., Tikekar, R.: Gravitation and Cosmology \textbf{11},
244 (2005)

\bibitem{JoMa}
John, A.J., Maharaj, S.D.:  Il Nuovo Cimento B \textbf{121}, 27
(2006)

\bibitem{MaTh}
Maharaj, S.D., Thirukkanesh, S.: Math. Meth. Appl. Sci.
\textbf{29}, 1943 (2006)

\bibitem{ThMa}
Thirukkanesh, S., Maharaj, S.D.: Class. Quantum Grav. \textbf{23},
2697 (2006)

\bibitem{DuBa}
 Durgapal, M.C., Bannerji, R.: Phys. Rev. D \textbf{27},
 328 (1983)

\bibitem{Tik}
Tikekar, R.: J. Math. Phys. \textbf{31}, 2454 (1990)

\bibitem{HaMa}
Hansraj, S., Maharaj, S.D.: Int. J. Mod. Phys. D \textbf{15}, 1311
(2006)

\bibitem{FiSk}
Finch, M.R., Skea, J.E.F.:  Class. Quantum Grav. \textbf{6}, 467
(1989)

\end{thebibliography}
\end{document}